# Quantum Materials: Shape Resonances in Superstripes

Antonio Bianconi

*Rome International Center for Materials Science Superstripes RICMASS, via dei Sabelli 119 A, 00185 Roma Italy*

*abstract:*

The significance of 'stripes' in certain high-temperature superconductors has been hotly debated for decades. Now a consensus is emerging that there may, in fact, be two networks of different stripes in which shape resonances play a key role in the superconductivity.

During the last week of May 2013 two hundred selected experts on unconventional superconductivity, quantum electronics and multiferroics gathered together for the international meeting, Quantum in Complex Matter, the fourth conference in the *Superstripes* series[1]. The site was the island of Ischia off the coast of Naples, Italy, which offers dramatic spots in a complex, unique landscape that has hosted Roman emperors Caligula and Tiberius, and writers such as Henrik Ibsen and Maxim Gorky. This series of conferences started in 1992 with the Phase Separation workshop organized by K. A. Müller in Erice, Italy, followed by the Stripes conferences held in Rome and Erice from 1996 to 2008, when the Superstripes conferences began. The focus has been on the mechanisms of high-temperature superconductivity controlled by lattice effects, phase separation and multi-condensate superconductivity[2]. These issues are at the forefront of research today after having been considered a 'minority paradigm' for 25 years.

During Superstripes 2013 the idea emerged that controlling high-temperature superconductivity by fine tuning the lattice features is not chemistry, but low-energy physics. Whereas chemistry involves interactions in the range of 1–6 eV, here multiple electronic components are tuned around the Fermi energy, in the range of 20–100 meV. This is close to the physics of living matter, where atoms in motion control the low-energy physics of interactions in the range of 25–200 meV.

Many hot topics were discussed, ranging from chiral excitations to spin excitations, and from the deviation of the local structure from the average structure to phase separation in iron-based superconductors and oxide interfaces, quantum electronics, pressure effects, fast dynamics, and resonant and inelastic X-ray diffraction studies. Here I focus on two subjects on which the community reached a solid consensus in high-temperature superconductivity.

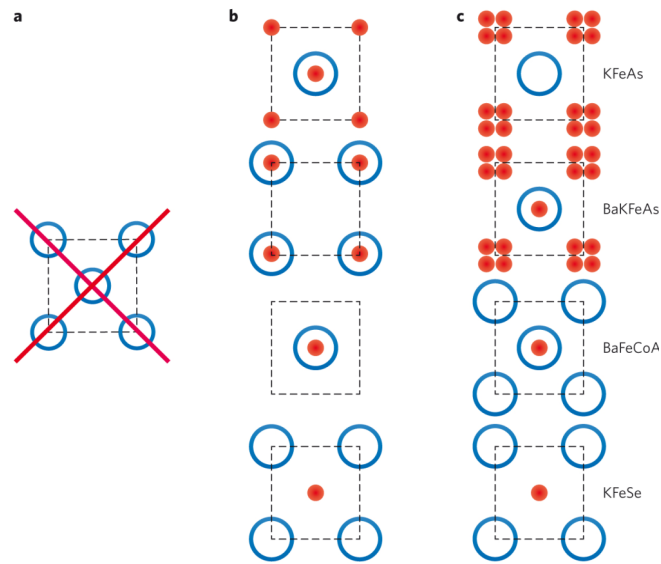

Figure 1 | Simplified Fermi surface topologies that favour multi-condensate superconductivity in iron-based superconductors. **a**, The $s\pm$ model. **b**, The shape resonance model. The blue circles are large Fermi surfaces and the red dots are small Fermi surface spots, which appear or disappear with small changes of the chemical potential at the centre or at the edges of the Brillouin zone (dashed line). **c**, Schematic of the Fermi surfaces detected by ARPES in different families of iron-based superconductors, as presented by Borisenko at Superstripes 2013[8,9]. All families show new Fermi surface spots at optimum $T_c$ that appear where the chemical potential is near a band edge, in agreement with the shape resonance model, whereas the nesting topology of the $s\pm$ model is never observed in the superconducting phase

As is well established, the Fermi topology should play a key role in discriminating between two proposed pairing mechanisms in iron-based superconductors: the $s\pm$ pairing[3] and the shape-resonance pairing[4–6]. Whereas the $s\pm$ model is based on nesting between two similar large Fermi surfaces, the shape-resonance idea is centred on the coexistence of at least one large Fermi surface and at least one small Fermi surface appearing or disappearing with small changes in the chemical potential (Fig. 1). The shape resonance in superconducting gaps[5] is a type of Fano–Feshbach resonance between pairing channels at a Bose–Einstein to Bardeen–Cooper–Schrieffer crossover in multiband superconductors, appearing when the chemical potential is tuned to the proximity of a band edge. Here the metal-to-metal 2.5 Lifshitz transition occurs because of the variation of the Fermi surface topology as a function of the chemical potential. By changing the chemical potential, the critical temperature ($T_c$) decreases towards 0 K when the chemical potential is tuned to the band edge — because of the Fano anti-resonance — and the $T_c$ maximum appears (as in Fano resonances) at higher energy, between one and two times the pairing interaction above the band edge. For iron-based superconductors comprising stacks of weakly connected atomic iron layers the maximum $Tc$ is predicted to occur where the energy dispersion along the $z$ direction is on the order of the energy of the pairing interaction.

The very recent high-resolution angle-resolved photoemission spectroscopy (ARPES) experiments presented by Atsushi Fujimori[7] on $BaFe_2(As_{1-x}P_x)_2$, and by Sergey Borisenko on many different iron-based superconductors[8,9] using variable photon energies, have succeeded in identifying the topology of the Fermi surface cylinders along the $z$ direction and the crossing of the Fermi level by the bands at the $\Gamma$ and Z points of the Brillouin zone. The compelling results show that similar hole-like and electron-like Fermi surfaces that are optimally nested (as proposed for the $s\pm$ mechanism) promote a magnetic phase. Contrary to this, high-temperature superconductivity appears for other Fermi surface topologies that are needed for the shape-resonance mechansim: these all have large Fermi surfaces of flat cylinders that coexist with vanishingly small Fermi surfaces where the band edge is close to the Fermi level. The doping moves the chemical potential by about 15–30 meV around the band edge. The dispersion of this band transverse to the iron plane is of the order of 20–30 meV with a pairing interaction of the order of 12–20 meV. The Fermi surface spot that appears has a large gap, whereas the larger Fermi surfaces have a smaller gap as predicted[5].

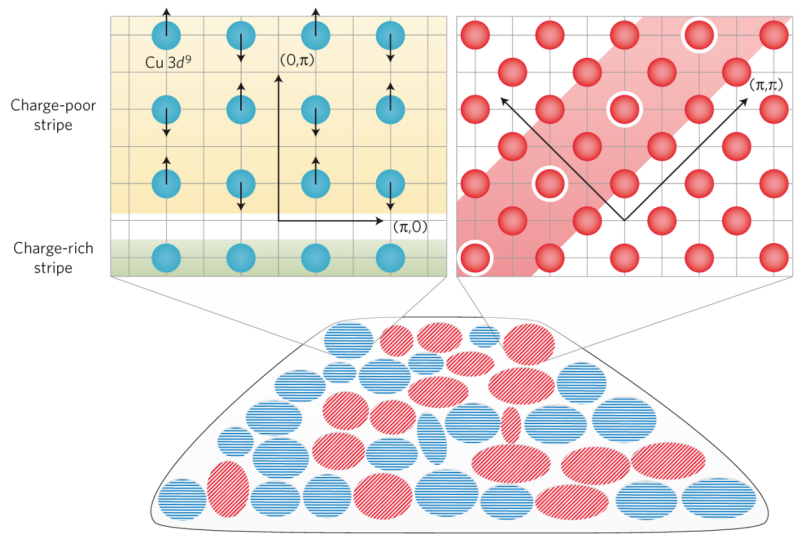

Figure 2 | Schematic of the superstripes scenario in the CuO2 plane, which make up the active atomic layers in the cuprate heterostructures. A complex phase separation forms networks of two different types of puddle, where blue indicates the magnetic puddles and red the superconducting puddles (bottom). Top left: Magnetic puddles comprise horizontal (or vertical) spin stripes in the Cu sublattice (blue dots), with spin indicated by black arrows. Top right: Superconducting puddles are made of diagonal lattice stripes in the oxygen sublattice (red dots) with rows of oxygen interstitials (white circles) in the spacer layers, and charge-rich stripes (red shading) intercalated by charge-poor stripes.

John Tranquada et al.10 and Barrett Wells et al.11 presented strong evidence that the spin stripes form nanoscale 'magnetic puddles' that are spatially separated from superconducting domains. Early results from Tranquada et al. on spin–charge stripes observed by inelastic neutron scattering in the cuprates12 were first presented at the Stripes and High $T$c Superconductivity conference held in Rome in 1996. The hole-rich rows are intercalated by three rows of hole-poor antiferromagnetic Cu lattice (Fig. 2) where the oxygen ions of the CuO2 lattice are removed. For many years, it has been difficult to reconcile these horizontal or vertical spin stripes with the diagonal lattice stripes13 in the oxygen ion sublattice shown in Fig. 2 due to the joint contribution of polaron self-organization13, lattice parameter misfit strain14 and self-organization of dopants in the spacer layers15. The topology of the diagonal lattice stripes are now unveiled using scanning nano X-ray diffraction, showing puddles that form a second scale-free network16. The lattice modulation in the striped puddles induces a Fermi surface reconstruction with mini-bands and partial gaps in the range of 20 meV where the Fermi level is near a band edge17, favouring the shape resonance in the striped puddles6.

These recent experiments lend support to the complex superstripes scenario18 of different networks of 'spin stripes' and 'lattice stripes' (Fig. 2), a nanoscale phase-separation similar to that in the pnictides19. Percolation and granular superconductivity20 will be important themes.

*Antonio Bianconi is at the Rome International Center for Materials Science, Superstripes, Via dei Sabelli 119/A, 00185 Roma, Italy.*
*e-mail: antonio.bianconi@ricmass.eu*